\documentstyle[twoside]{article}

\catcode`\@=11
\long\def\@makefntext#1{
\protect\noindent \hbox to 3.2pt {\hskip-.9pt

$^{{\eightrm\@thefnmark}}$\hfil}#1\hfill}               

\def\thefootnote{\fnsymbol{footnote}}
\def\@makefnmark{\hbox to 0pt{$^{\@thefnmark}$\hss}}    

\def\ps@myheadings{\let\@mkboth\@gobbletwo
\def\@oddhead{\hbox{}
\rightmark\hfil\eightrm\thepage}

\def\@oddfoot{}\def\@evenhead{\eightrm\thepage\hfil
\leftmark\hbox{}}\def\@evenfoot{}
\def\sectionmark##1{}\def\subsectionmark##1{}}

\oddsidemargin=\evensidemargin
\renewcommand{\thefootnote}{\fnsymbol{footnote}}

\newcounter{sectionc}\newcounter{subsectionc}\newcounter{subsubsectionc}
\renewcommand{\section}[1] {\vspace{12pt}\addtocounter{sectionc}{1}

\setcounter{subsectionc}{0}\setcounter{subsubsectionc}{0}\noindent

        {\tenbf\thesectionc. #1}\par\vspace{5pt}}
\renewcommand{\subsection}[1] {\vspace{12pt}\addtocounter{subsectionc}{1}

        \setcounter{subsubsectionc}{0}\noindent

        {\bf\thesectionc.\thesubsectionc. {\kern1pt \bfit #1}}\par\vspace{5pt}}
\renewcommand{\subsubsection}[1] {\vspace{12pt}\addtocounter{subsubsectionc}{1}
        \noindent{\tenrm\thesectionc.\thesubsectionc.\thesubsubsectionc.
        {\kern1pt \tenit #1}}\par\vspace{5pt}}
\newcommand{\nonumsection}[1] {\vspace{12pt}\noindent{\tenbf #1}
        \par\vspace{5pt}}

\newcounter{appendixc}
\newcounter{subappendixc}[appendixc]
\newcounter{subsubappendixc}[subappendixc]
\renewcommand{\thesubappendixc}{\Alph{appendixc}.\arabic{subappendixc}}
\renewcommand{\thesubsubappendixc}
        {\Alph{appendixc}.\arabic{subappendixc}.\arabic{subsubappendixc}}

\renewcommand{\appendix}[1] {\vspace{12pt}
        \refstepcounter{appendixc}
        \setcounter{figure}{0}
        \setcounter{table}{0}
        \setcounter{lemma}{0}
        \setcounter{theorem}{0}
        \setcounter{corollary}{0}
        \setcounter{definition}{0}
        \setcounter{equation}{0}
        \renewcommand{\thefigure}{\Alph{appendixc}.\arabic{figure}}
        \renewcommand{\thetable}{\Alph{appendixc}.\arabic{table}}
        \renewcommand{\theappendixc}{\Alph{appendixc}}
        \renewcommand{\thelemma}{\Alph{appendixc}.\arabic{lemma}}
        \renewcommand{\thetheorem}{\Alph{appendixc}.\arabic{theorem}}
        \renewcommand{\thedefinition}{\Alph{appendixc}.\arabic{definition}}
        \renewcommand{\thecorollary}{\Alph{appendixc}.\arabic{corollary}}
        \renewcommand{\theequation}{\Alph{appendixc}.\arabic{equation}}
        \noindent{\tenbf Appendix \theappendixc #1}\par\vspace{5pt}}
\newcommand{\subappendix}[1] {\vspace{12pt}
        \refstepcounter{subappendixc}
        \noindent{\bf Appendix \thesubappendixc. {\kern1pt \bfit #1}}
        \par\vspace{5pt}}
\newcommand{\subsubappendix}[1] {\vspace{12pt}
        \refstepcounter{subsubappendixc}
        \noindent{\rm Appendix \thesubsubappendixc. {\kern1pt \tenit #1}}
        \par\vspace{5pt}}

\newcommand{\textlineskip}{\baselineskip=13pt}
\newcommand{\smalllineskip}{\baselineskip=10pt}

\def\eightcirc{
\begin{picture}(0,0)
\put(4.4,1.8){\circle{6.5}}
\end{picture}}
\def\eightcopyright{\eightcirc\kern2.7pt\hbox{\eightrm c}}

\newcommand{\copyrightheading}[1]
        {\vspace*{-2.5cm}\smalllineskip{\flushleft
        {\footnotesize International Journal of Modern Physics A, #1}\\
        {\footnotesize $\eightcopyright$\, World Scientific Publishing
         Company}\\
         }}

\def\abstracts#1#2#3{{
        \centering{\begin{minipage}{4.5in}\baselineskip=10pt\footnotesize
        \parindent=0pt #1\par

        \parindent=15pt #2\par
        \parindent=15pt #3
        \end{minipage}}\par}}

\newcommand{\bibit}{\nineit}

\renewenvironment{thebibliography}[1]
        {\frenchspacing
         \ninerm\baselineskip=11pt
         \begin{list}{\arabic{enumi}.}
        {\usecounter{enumi}\setlength{\parsep}{0pt}
         \setlength{\leftmargin 12.7pt}{\rightmargin 0pt} 
         \setlength{\itemsep}{0pt} \settowidth
        {\labelwidth}{#1.}\sloppy}}{\end{list}}

\newcounter{itemlistc}
\newcounter{romanlistc}
\newcounter{alphlistc}
\newcounter{arabiclistc}
\newenvironment{itemlist}
        {\setcounter{itemlistc}{0}
         \begin{list}{$\bullet$}
        {\usecounter{itemlistc}
         \setlength{\parsep}{0pt}
         \setlength{\itemsep}{0pt}}}{\end{list}}

\newenvironment{romanlist}
        {\setcounter{romanlistc}{0}
         \begin{list}{$($\roman{romanlistc}$)$}
        {\usecounter{romanlistc}
         \setlength{\parsep}{0pt}
         \setlength{\itemsep}{0pt}}}{\end{list}}

\newcommand{\fcaption}[1]{
        \refstepcounter{figure}
        \setbox\@tempboxa = \hbox{\footnotesize Fig.~\thefigure. #1}
        \ifdim \wd\@tempboxa > 5in
           {\begin{center}
        \parbox{5in}{\footnotesize\smalllineskip Fig.~\thefigure. #1}
            \end{center}}
        \else
             {\begin{center}
             {\footnotesize Fig.~\thefigure. #1}
              \end{center}}
        \fi}

\newcommand{\tcaption}[1]{
        \refstepcounter{table}
        \setbox\@tempboxa = \hbox{\footnotesize Table~\thetable. #1}
        \ifdim \wd\@tempboxa > 5in
           {\begin{center}
        \parbox{5in}{\footnotesize\smalllineskip Table~\thetable. #1}
            \end{center}}
        \else
             {\begin{center}
             {\footnotesize Table~\thetable. #1}
              \end{center}}
        \fi}

\def\@citex[#1]#2{\if@filesw\immediate\write\@auxout
        {\string\citation{#2}}\fi
\def\@citea{}\@cite{\@for\@citeb:=#2\do
        {\@citea\def\@citea{,}\@ifundefined
        {b@\@citeb}{{\bf ?}\@warning
        {Citation `\@citeb' on page \thepage \space undefined}}
        {\csname b@\@citeb\endcsname}}}{#1}}

\newif\if@cghi
\def\cite{\@cghitrue\@ifnextchar [{\@tempswatrue
        \@citex}{\@tempswafalse\@citex[]}}
\def\citelow{\@cghifalse\@ifnextchar [{\@tempswatrue
        \@citex}{\@tempswafalse\@citex[]}}
\def\@cite#1#2{{$\null^{#1}$\if@tempswa\typeout
        {IJCGA warning: optional citation argument

        ignored: `#2'} \fi}}

\def\pmb#1{\setbox0=\hbox{#1}
        \kern-.025em\copy0\kern-\wd0
        \kern.05em\copy0\kern-\wd0
        \kern-.025em\raise.0433em\box0}

\def\fnt#1#2{\footnotetext{\kern-.3em
        {$^{\mbox{\scriptsize #1}}$}{#2}}}

\def\fpage#1{\begingroup
\voffset=.3in
\thispagestyle{empty}\begin{table}[b]\centerline{\footnotesize #1}
        \end{table}\endgroup}

\font\tenrm=cmr10
\font\tenit=cmti10

\font\tenbf=cmbx10
\font\bfit=cmbxti10 at 10pt
\font\ninerm=cmr9
\font\nineit=cmti9

\font\eightrm=cmr8

\def\qed{\hbox{${\vcenter{\vbox{                        
   \hrule height 0.4pt\hbox{\vrule width 0.4pt height 6pt
   \kern5pt\vrule width 0.4pt}\hrule height 0.4pt}}}$}}

\renewcommand{\thefootnote}{\fnsymbol{footnote}}        

\begin{document}


\normalsize\textlineskip
\thispagestyle{empty}
\setcounter{page}{1}

\copyrightheading{}                     

\vspace*{0.88truein}

\fpage{1}
\centerline{\bf 1+1 DIMENSIONAL YANG-MILLS THEORIES}
\vspace*{0.035truein}
\centerline{\bf IN LIGHT-CONE GAUGE\footnote{Invited report at the Workshop
``Low Dimensional Field Theory'', Telluride (CO), Aug. 5-17, 1996}}
\vspace*{0.37truein}
\centerline{\footnotesize A. BASSETTO}
\vspace*{0.015truein}
\centerline{\footnotesize\it Dipartimento di Fisica and INFN Sezione di Padova,
 }
\baselineskip=10pt
\centerline{\footnotesize\it Universit\'a di Padova, via Marzolo 8, 35131
Padova, Italy}
\vspace*{10pt}
\centerline{\footnotesize G. NARDELLI}
\vspace*{0.015truein}
\centerline{\footnotesize\it Dipartimento di Fisica and INFN Gruppo collegato
di Trento} \baselineskip=10pt
\centerline{\footnotesize\it Universit\'a di Trento, 38050 Povo (TN), Italy}
\vspace*{0.225truein}

\vspace*{0.21truein}
\abstracts{In 1+1 dimensions two different formulations exist
of SU(N) Yang Mills theories in light-cone gauge; only one of them
gives results which comply with the ones obtained in Feynman
gauge. Moreover the theory, when considered in 1+(D-1) dimensions,
looks discontinuous  in the limit D=2. All those
features are proven in Wilson loop calculations as well as in the
study of the $q\bar q$ bound state integral equation in the
large N limit.}{}{}


\vspace*{1pt}\textlineskip      
\section{Introduction and motivations}    
\vspace*{-0.5pt}
\noindent
In this report, we review some properties concerning Yang-Mills (YM)
theories in 1+1 dimensions in the light-cone gauge. The reason why YM
theories in 1+1 dimensions are interesting is at least twofold:
\begin{itemlist}
 \item The reduction of the dimensions to $D=2$ entails tremendous
simplifications in the theory, so that several important problems can be faced
in this lower dimensional context. We are thinking  for instance to the exact
(when possible) evaluation of vacuum to vacuum amplitudes of Wilson loop
operators, that, for a suitable choice of contour and in some specific limit,
provide the potential between two static quarks. Another example is the
spectrum of the Bethe-Salpeter equation, when dynamical fermions are added to
the system.
 \item The second reason is that YM theories in $D=2$ have several peculiar
features that are interesting by their own. The most remarkable ones are:

a) in $D=2$ within the same gauge choice (light-cone gauge) two
inequivalent formulations of the theory seem to coexist;

b) $D=2$ is a point of discontinuity for YM theories;
this is an intriguing
feature whose meaning has not been fully understood so far.
\end{itemlist}

\noindent
All the features we have listed are most conveniently studied
if the light-cone
gauge (lcg) is chosen. In such a gauge the Faddeev Popov sector
decouples and the unphysical degrees of freedom content of the theory is
minimal. The price to be paid for these nice features is the presence of the so
called `spurious' poles in the vector propagator.\cite{BNS91}
In fact, in the gauge $nA=0$ with $n^\mu$ a given constant null vector
($n^2=0$), the form of the
propagator in $D$ dimensions turns out to be
\begin{equation}
D^{ab}_{\mu \nu} (k)= {-i \delta^{ab}\over k^2 + i \epsilon} \left( g_{\mu \nu}
- {n_\mu k_\nu + n_\nu k_\mu\over nk}\right)\, .
 \label{prop1}
\end{equation}
As we shall see, to handle the spurious pole at $nk=0$  is a
delicate matter; basically all difficulties encountered in the past
within the lcg quantization are related to this problem.

In Sect. 2 we  focus on the $D\ne 2$ case, and  discuss the so
called `manifestly unitary' and `causal' formulations of the theory. We shall
see that the correct formulation is the causal one: the
manifestly unitary formulation will meet so many inconsistencies to make it
unacceptable. Moreover, even in the causal formulation, the theory looks
discontinuous in the limit $D=2$.

In Sect. 3 we  compare the two formulations at strictly $D=2$. Surprisingly, in
this case both seem to coexist, without obvious inconsistencies.
Thus, a natural question arises: are the two quantization schemes equivalent in
$D=2$? Do they provide us with equal results?

The
answer to these questions are given in Sect. 4 where Wilson loop
expectation values are evaluated. We shall find
that the two formulations are indeed inequivalent.

In Sect. 5
the theory will be considered on the cylinder ${\cal R}\times {\cal S}$,
namely with the space variable constrained in an interval, in order to
reach a consistent infrared (IR) regularization. Again the two formulations
behave quite differently.

Finally Sect. 6 contains a discussion
of the bound state integral equation when dynamical fermions are present
and our conclusions.

\textheight=7.8truein
\setcounter{footnote}{0}
\renewcommand{\thefootnote}{\alph{footnote}}

\section{$D\ne 2$: a comparison between manifestly unitary and causal
formulations}
\noindent
A manifestly unitary formulation of YM theories in lcg can be obtained by
quantizing the theory in the so called null-frame formalism, i.e. passing in
light-cone coordinates and interpreting $x^+$ as the evolution coordinate
(time) of the system; the remaining components $x^-, x_\perp$ will be
interpreted as  `space' coordinates. Within this quantization scheme, one of
the unphysical components of the gauge potential (say $A_-$) is set equal to
zero by the gauge choice whereas the remaining unphysical component ($A_+$)
is no
longer a dynamical variable but  rather a Lagrange multiplier of the secondary
constraint (Gauss' law). Thus, already at the classical level, it is possible
to restrict to the phase space containing only the physical (transverse)
polarization of the gauge fields. Then, canonical quantization on the null
plane provides the answer to the prescription for the spurious pole in the
propagator, the answer being essentially the Cauchy principal value (CPV)
prescription.

Unfortunately, following this scheme, several
inconsistencies arise, all of them being related to the violation of causality
that $CPV$ prescription entails:
 \begin{itemlist}
\item non-renormalizability of the theory: already at the one loop level,
 dimensionally regularized Feynman integrals develop loop singularities that
appear as double poles  at $D=4$.\cite{CDL85}
\item power counting criterion is lost: the pole structure in the complex $k_0$
plane is such that spurious poles contribute under Wick rotation. As a
consequence euclidean Feynman integrals are not
simply related to Minkowskian ones as an extra contribution shows up
which jeopardizes naive power counting.\cite{BW89}
\item gauge invariance is lost: due to the above mentioned extra
contributions, the $N=4$ supersymmetric version of the theory turns out
not to be finite, at variance with the Feynman gauge result.\cite{CDL85}
\end{itemlist}

Consequently, manifestly unitary theories do not seem to exist. As explained
above, all the bad features of this formulation have their root in the lack of
causality of the prescription for the spurious pole, and the subsequent
failure of the power counting criterion for convergence.
Thus, a natural way to circumvent
these problems is to choose a causal prescription. It was
precisely following these arguments  that Mandelstam and
Leibbrandt \cite{ML83}, independently, introduced  the ML prescription
\begin{equation}
{1\over k_-}\equiv ML({1\over k_-})= {k_+\over k_+k_- + i \epsilon}={1\over
k_- +
i \epsilon {\rm sign}(k_+)}\, .
\label{ml}
\end{equation}
It can be easily realized that with this choice the position of
the spurious pole is always `coherent' with that of Feynman ones, no
extra terms appearing after Wick rotation which threaten the power counting
criterion for convergence.
How can one justify such a recipe? One year later Bassetto and
collaborators \cite{BDLS85} filled the gap
by showing that ML prescription arises naturally by quantizing the
theory at equal time,
rather than at equal $x^+$. Eventually they succeeded \cite{BDS87} in proving
full renormalizability of
the theory and full agreement with Feynman gauge results in
perturbative
calculations.\cite{BKKN93}

At present the level of
accuracy of the light-cone gauge is indeed comparable with that of the
covariant gauges.

An important point to be stressed is that equal time canonical quantization
in lcg,
leading to the ML prescription for the spurious pole, does not provide us
with a manifestly
unitary formulation of the theory. In fact in this formalism Gauss' laws do not
hold strongly but, rather, the Gauss' operators obey to a free field equation
and entail the presence in the Fock
space of unphysical degrees of freedom. The causal
nature of the ML prescription for the spurious poles is a consequence of
the causal propagation of those `ghosts'. A physical
Hilbert space can be selected by imposing the (weakly) vanishing
of Gauss' operators.  This mechanism is similar to the Gupta Bleuler
quantization scheme for electrodynamics in Feynman gauge,
but with the great advantage that it can be naturally extended to the non
abelian case without Faddeev Popov ghosts.\cite{BNS91}
\vfill\eject

\textheight=7.8truein
\setcounter{footnote}{0}
\renewcommand{\thefootnote}{\alph{footnote}}

\section{$D= 2$: a comparison between the manifestly unitary and causal
formulations}
\noindent
The causal formulation of the theory can be straightforwardly extended to {\it
any} dimension, including the case $D=2$. On the other hand, the
manifestly unitary formulation can {\it only} be defined in $D=2$
without encountering obvious inconsistencies. The reason is simple:
all problems
were related to the lack of causality encoded in the $CPV$ prescription.
But at exactly $D=2$ there are no physical degrees of
freedom propagating at all, and then causality is no longer a concern.
Moreover, at exactly $D=2$ and within the lcg, the 3- and 4- gluon vertices
vanish, so that all the inconsistencies related to the perturbative evaluation
of Feynman integrals are no longer present in this case.
A manifestly unitary formulation provides the following
`instantaneous - Coulomb type' form for the only non vanishing component of the
propagator:
\begin{equation}
D^{ab}_{++} (x) = - {i\delta^{ab}\over (2\pi)^2}\int d^2 k \, e^{ikx}
{\partial\over \partial k_-} P\left({1\over k_-}\right) = -i\delta^{ab}
{|x^-|\over 2} \delta(x^+)\, , \label{prcpv2}
\end{equation}
where $P$ denotes CPV prescription,
whereas equal time canonical quantization gives, for the same component of the
propagator,
\begin{equation}
D_{++}^{ab}(x)= {i\delta^{ab}\over \pi^2}\int d^2k\, e^{ikx} {k_+^2\over
(k^2+i\epsilon)^2}={\delta^{ab}(x^-)^2\over \pi ( -x^2 + i \epsilon)}\ .
\label{prml2}
\end{equation}
Thus, it seems we have  two different formulation of YM theories in
$D=2$, and within the same gauge choice, the lcg.\cite{BDG94}
Whether they are equivalent and, in
turn, whether they are equivalent to a different gauge choice, such as
Feynman gauge,  has to be explicitly verified.

We can summarize the situation according to the content of unphysical degrees
of freedom. Since the paper by 't Hooft in 1974 \cite{TH74},
it is a common belief that
pure YM in
$D=2$ is a theory with no propagating degrees of freedom. This
happens in the
manifestly unitary formulation leading to CPV prescription for the spurious
pole and to the propagator (\ref{prcpv2}).
This formulation, however, cannot be
extended outside $D=2$ without inconsistencies. Alternatively, we have the same
gauge choice but with a different quantization scheme, namely at equal time,
leading to the
causal (ML) prescription for the spurious pole and to the propagator
(\ref{prml2}). Here, even in the pure YM case, some
degrees of freedom survive, as we have propagating ghosts.
Such a formulation is in a better shape when compared to the
previous one as it can be  smoothly extended to any dimension,
where consistency
with Feynman gauge has been established.

Feynman gauge validity for
any $D\ne 2$ is unquestionable, while, at strictly $D=2$, the vector
propagator in this gauge
fails to be a tempered distribution. Still, in the spirit of dimensional
regularization, one can always evaluate amplitudes in $D\ne 2$ and take
eventually the limit $D\to 2$. In following this attitude, the number of
degrees of freedom
is even bigger as Faddeev-Popov ghosts are also to be taken into account.
In addition,
in the covariant gauge 3- and 4- gluon vertices do not
vanish and the theory does not look free at all.

\textheight=7.8truein
\setcounter{footnote}{0}
\renewcommand{\thefootnote}{\alph{footnote}}

\section{Wilson loops calculations}
\noindent
To clarify the whole matter we need a test of gauge invariance. In particular,
we want to answer the following three questions:
\begin{romanlist}
\item Is YM theory continuous in the limit $D\to 2$?
\item Is YM theory in $D=2$ a free theory?
\item Are the two lcg formulations in $D=2$ equivalent?
\end{romanlist}
\noindent
To probe gauge invariance and to answer the above questions, following
ref.\cite{BDG94}, we shall evaluate
vacuum to vacuum  amplitudes of  Wilson loop operators, defined as a
functional of the closed contour $\gamma$ through
\begin{equation}
W[\gamma]= {1\over N} \int dA_\mu \delta (\Phi (A)) {\rm det} [M_\Phi] e^{i\int
dx {\cal L}(x)}\, {\rm Tr} \left\{ {\cal P} e^{i g \oint_\gamma dx^\mu A^a_\mu
T^a}\right\}\, ,\label{wl}
\end{equation}
where for convenience we choose $SU(N)$ as gauge group with hermitean
generators $T^a$. In Eq.(\ref{wl}), $\Phi(A)=0$ denotes the gauge choice and
${\rm det}[M_\Phi]$ the corresponding Faddeev-Popov determinant, that can be
either trivial or not, depending on the gauge $\Phi(A)$. As usual,
${\cal P}$ denotes ordering along the closed path $\gamma$, that we shall
choose to be  a light-like rectangle in the plane  $(x^+, x^-)$
 with length sides $(T,L)$.
For later convenience, we recall that the Casimir constants of the
fundamental and adjoint representations, $C_F$ and $C_A$, are defined through
\begin{equation}  C_F= (1/N) {\rm Tr} (T^a T^a)= (N^2-1)/2N\ , \quad {\rm and}
\quad C_A \delta^{ab} = f^{acd}f^{bcd}= \delta^{ab} N \ , \label{casimir}
\end{equation}
$f^{abc}$ being the structure constants for $SU(N)$.

First of all we shall check continuity in the $D\to 2$ limit. To this
purpose, we have to choose the lcg in its causal formulation. In fact,
among the
gauge choices we considered, this is the only one whose formulation is
smooth in the $D\to 2$ limit\footnote{In fact, lcg in its manifestly unitary
formulation is acceptable only at $D=2$ and therefore cannot be used to check
continuity; even Feynman gauge cannot be used, as the propagator is divergent
at $D=2$, preventing therefore a calculation at exactly $D=2$.}.
Within this gauge choice, only a perturbative ${\cal O}(g^4)$ calculation is
viable. Performing the calculation in $D$ dimensions and
eventually taking the limit $D\to 2$, the expression for the Wilson loop gives
\begin{equation}
\lim_{D\to2} W_{ML}^{(D)}(\gamma) = 1-i {g^2\over 2}{LTC_F} -{g^4\over 8}(LT)^2
\left[C^2_F-{C_FC_A\over 8\pi^2}\left(1+{\pi^2\over 3}\right)\right] + {\cal
O}(g^6)\ ,\label{wlmld}
\end{equation}
whereas the same quantity evaluated at exactly $D=2$ gives a different answer,
namely
\begin{equation}
 W_{ML}^{(D=2)}(\gamma) = 1-i {g^2\over 2}{LTC_F} -{g^4\over 8}(LT)^2
\left[C^2_F-{C_FC_A\over 24}\right] + {\cal
O}(g^6)\ .\label{wlml2}
\end{equation}
Thus, we have a surprising result: YM theories are discontinuous at $D=2$.
The technical reason of such discontinuity can be easily understood
in terms of
`anomalous' diagrams that survive in the limit $D\to 2$. In strictly $D=2$, as
already stressed, the 3- and 4- gluon vertices vanish in lcg.
Consequently, the free propagator (\ref{prml2}) is the $complete$ two
point Green
function, as there are no radiative corrections. On the other hand, in
$D=2+\varepsilon$ the gluon vertices do not vanish anymore, as `$\varepsilon$'
transverse components couple the gauge fields. Thus, in $D\ne 2$ dimensions
the two-point Green function has radiative corrections. The one loop
correction, ${\cal O}(g^2)$, is the standard `bubble diagram', with two free
propagators connected by two 3-gluon vertices. Obviously, the strength of the
vertices vanishes in the limit $\varepsilon=(D-2) \to 0$; nevertheless,this
correction to the Green function produce a finite contribution in the limit
$\varepsilon \to 0$ due to the matching with the loop pole precisely at
$D=2$. Such a dimensional `anomaly-type' phenomenon is responsible of the
discontinuity of YM theory at $D=2$.

As a matter of fact, it is easy to
evaluate the contribution to the Wilson loop given by this anomalous part of
Green function, surviving in the limit $D\to 2$:  it provides the factor $g^4
(LT)^2 C_FC_A/64\pi^2$, which is indeed the difference between Eq.~(\ref{wlmld})
and Eq.~(\ref{wlml2}). This discontinuity at $D=2$ is a very interesting
phenomenon, whose nature is still unclear; whether it is related to a
true anomaly, $i.e.$ whether there is a classical symmetry violated
at the quantum level,
is still a matter of investigation.

Consistency with Feynman gauge can be checked by evaluating the
dimensionally regularized Wilson loop: an ${\cal O}(g^4)$
calculation provides exactly the same result of lcg in its causal formulation
for any $D$ and therefore also in the limit $D\to 2$: as expected, we have full
agreement between Feynman and light-cone gauge if the ML prescription for the
spurious poles is adopted. We stress that the `anomalous' self-energy
contribution we have hitherto discussed,
is essential in order to get such an agreement.\cite{BKKN93}

However, both in the dimensionally regularized case
with the limit $D\to 2$ taken at the end and in the strictly $D=2$ case,
within the causal lcg formulation, we realize that the Wilson loop
results do not depend only on $C_F$:
a `genuine' non abelian $C_FC_A$ dependence appears at ${\cal O}(g^4)$.
This means that, although the vertices vanish, YM in $D=2$ dimensions is
$not$ equivalent to an abelian theory.

On the contrary, this feature does not occur in the manifestly unitary
(strictly 2-dimensional) formulation. In fact, due to the contact nature
of the propagator, Eq.~(\ref{prcpv2}), non-abelian $C_A$-dependent  terms
do not appear in the expression of Wilson loops. In this case, it is easy to
find that the perturbative result exponentiates in a simple abelian way
\begin{equation}
W_{CPV}^{(D=2)}(\gamma)= e^{-ig^2 LT C_F/2}\, .
\label{wlcpv2}
\end{equation}
Pure YM in its manifestly unitary formulation is essentially free and
equivalent
to an abelian theory. We are lead to conclude that the
two light-cone formulations at $D=2$ are indeed inequivalent.

Summarizing, three different evaluations of the same Wilson loop within the
same gauge choice (lcg) provided us with three different
answers! Discrepancy between Eqs. (\ref{wlml2}) and (\ref{wlcpv2})
is explained by the coexistence of two different  inequivalent formulations of
YM theory, whereas discrepancy between Eqs.
(\ref{wlmld}) and (\ref{wlml2}) is explained by the discontinuity in
the limit $D\to 2$.

However, in all the cases we considered, we always got  at least
a pure `area-law' dependence of the Wilson loop.
Is this an universal property of  $D=2$ YM theory? Contrary to a
common belief, we shall show that this is not the case, by providing an
explicit counterexample.\cite{BCN96}

Let us consider again a  rectangular
loop $\tilde \gamma$ with area $A=LT$, but now centered at the origin of
the plane  $(x^0,x^1)$. For convenience, let us stick to the $D=2$ case
and focus on the two different formulations of lcg. From a
physical point of view, this contour is even more interesting: would one be
able to compute the exact value of the Wilson loop amplitude, one could
derive
the potential $V(L)$ between two static quarks separated by a distance $L$
through the well known formula
\begin{equation}
\lim_{T\to \infty} W(\tilde \gamma)= e^{-i T V(L)}
\label{potential}
\end{equation}
In the manifestly unitary (CPV) case, due to the contact nature of the
potential (\ref{prcpv2}), the Wilson loop can again be exactly evaluated
giving, for a finite size of the rectangle,
\begin{equation}
W_{CPV}^{(D=2)}(\tilde \gamma) = e^{-ig^2 C_F LT/2}\
\label{loop2cpv}
\end{equation}
and therefrom a linear confining potential between quarks with string tension
$\sigma= g^2 C_F/2$. However, it should be emphasized that such a confining
result for $QCD_2$ has the same origin as in QED, namely follows from
the $abelian$ `contact' nature of the potential.
In the causal (ML) case a complete evaluation at all orders is not
viable due to the presence of genuine non abelian terms. Only a
perturbative ${\cal O}(g^4)$
evaluation is viable and after lengthy calculations, one finds
\begin{eqnarray}
W_{ML}^{(D=2)}(\tilde\gamma)&=& 1-i {g^2\over 2}{LTC_F} -{g^4\over 8}(LT)^2
\left\{C_F^2 + {C_FC_A\over 4\pi^2}\left[ 3 + {2\pi^2\over 3} + 2\beta[1+
\right. \right.\nonumber\\
&&\left.\left. (2+\beta)\ln\beta] - 2 (1+\beta)^2 \ln(1+\beta) - {2\over 3\beta}
\ln^2 (1+\beta) - {1\over 6\beta^2} \times
\right.\right.\nonumber\\
&&\left.\left.(1-\beta)^2\ln^2 (1-b) -{1\over
3\beta^2}(1-\beta)^4 \left({\rm Li}(\beta)+{\rm Li}\left(-{\beta\over
1+\beta}\right)\right)- \right.\right.\nonumber\\
 &&\left.\left. {1\over \beta} \left({\rm Li} (\beta) + {\rm Li}
\left({\beta\over 1 + \beta}\right)\right)\right]\right\} + {\cal O} (g^6)\ .
\label{loop2ml}
\end{eqnarray}
The Wilson loop amplitude, for finite $L$ and $T$, not only depends on
the area,  but also on the dimensionless ratio $\beta = L/T$ through a
complicated factor involving the dilogarithm function ${\rm Li} (z)$.

Obviously,
the fact that in this case we only have a perturbative ${\cal O}(g^4)$
calculation prevent us from making any interpretation of the result in term of
a potential between static quarks in the large $T$ limit. Nevertheless, it is
remarkable and perhaps not incidental that in such a limit
all the dependence on
$\beta$ cancel leaving again a pure area dependence
\begin{equation}
\lim_{T\to\infty}W_{ML}^{(D=2)}(\tilde\gamma)= 1-i {g^2\over 2}{LTC_F}
-{g^4\over
8}(LT)^2 \left\{C_F^2 + {C_FC_A\over 12\pi^2}(9 + 2\pi^2) \right\}\
\label{loop2mllarget}
\end{equation}
with finite coefficients.
We stress that again the same theory with the same gauge choice leads to
{\it different} results when using different expressions for the two
point Green function, even in the large T limit.

\section{Wilson loops on the cylinder}
\noindent
While a comparison with Feynman gauge at $D\ne2$ gave a satisfactory result,
a comparison at strictly $D=2$ is impossible owing to the well-known IR
singular behaviour of the vector propagator in Feynman gauge. Then, in
order to achieve a consistent IR regularization, we consider the theory
on the cylinder ${\cal R}\times {\cal S}$, namely we restrict the space variable
to the interval $-L\le x \le L$ with periodic boundary conditions on the
potentials. Time is $not$ compactified. We follow here the treatment given
in ref.\cite{BGN96}.

In so doing new features appear owing to the non trivial topology of the
cylinder and we feel preliminary to examine the equal-time quantization
of the pure YM theory in the light-cone gauge $A_{-}=0$. Introduction
of fermions at this stage would not entail particular difficulties,
but would be inessential to our subsequent argument.

We recall that axial-type gauges cannot be defined on compact manifolds
without introducing singularities in the vector potentials (Singer's
theorem).\cite{SI78} Partial compactifications are possible provided they occur
in a direction different from the one of the gauge fixing vector: this
is indeed what happens in the present case.

Starting from the standard lagrangian density (for SU(N))
\begin{equation}
{\cal L}= -1/2\, Tr(F^{\mu\nu}F_{\mu\nu})\, - 2 Tr(\lambda nA),
\label{lagrangian}
\end{equation}
$n_{\mu}={1\over \sqrt{2}}(1,1)$ being the gauge vector and $\lambda$
being Lagrange multipliers, which actually coincide with Gauss' operators,
it is straightforward to derive
the equations of motion
\begin{eqnarray}
&A_{-}=0,\,\,\,{\partial_{-}}^2 A_{+}=0,\nonumber\\
&\partial_{-}\partial_{+}A{+} -ig[A_{+},\partial_{-}A_{+}]=\lambda.
\label{motion}
\end{eqnarray}
As a consequence we get
\begin{equation}
\label{gausson}
\partial_{-}\lambda=0.
\end{equation}
In a `light-front' treatment (quantization at equal $x^{+}$), this
equation would be a constraint and $\partial_{-}$ might be inverted
(with suitable boundary conditions) to get the `strong' Gauss' laws
\begin{equation}
\label{gauss}
\lambda=0.
\end{equation}
This would correspond in the continuum to the $CPV$ prescription for
the singularity at $k_{-}=0$ in the relevant Green functions.

In equal-time quantization eq.(\ref{gausson}) is an evolution equation. The
Gauss' operators do not vanish strongly: Gauss' laws are imposed
as conditions on the `physical' states of the theory.
In so doing one can show \cite{BGN96} that the only surviving `physical'
degrees of freedom are zero modes of the potentials related
to phase factors of contours winding around the cylinder.
Frequency parts are unphysical, but non vanishing: they contribute
indeed to the causal expression of the vector propagator
\begin{eqnarray}
G_{c}(t,x)&=& G(t,x) -{{it}\over {4L}}P ctg \big({{\pi\sqrt2 x^{+}}
\over {2L}}\big),\nonumber\\
G(t,x)&=&1/2\, |t|\, \big(\delta_{p}(x+t) - {1\over {2L}} \big),
\label{propa}
\end{eqnarray}
$\delta_{p}$ being the periodic generalization of the Dirac distribution.

$G_{c}$ looks like a complex ``potential" kernel, the absorptive part being
related to the presence of ghost-like excitations, which are
essential to recover the ML prescription in the decompactification
limit $L \to \infty$; as a matter of fact
in this limit $G(t,x)$ becomes the
`instantaneous'  't Hooft potential, whereas $G_{c}$ is turned into the
causal ML distribution.

We are now in the position of comparing a Wilson loop on the cylinder
when evaluated according to the 't Hooft potential or using the
causal light-cone propagator.

In order to avoid an immediate interplay with topological features, we
consider a Wilson loop entirely contained in the basic interval
$-L\le x \le L$. We choose again a rectangular
Wilson loop $\gamma$ with light--like sides, directed along the vectors
$n_\mu$ and $n^*_\mu$, with lengths $\lambda$ and $\tau$
respectively, and parametrized according to the equations:

\begin{eqnarray}
\label{quarantasei}
C_1:x^\mu (s) &=& n^{\mu} {\lambda} s, \nonumber\\
C_2:x^\mu (s) &=& n^{\mu} {\lambda}+ n^{* \mu}{\tau} s, \nonumber\\
C_3:x^\mu (s) &=& n^{* \mu}{\tau} + n^{\mu} {\lambda}( 1-s), \\
C_4:x^\mu (s) &=& n^{* \mu}{\tau} (1 - s), \qquad 0 \leq s \leq 1, \nonumber
\end{eqnarray}
with ${\lambda + \tau}<2\sqrt 2 L$.
We are again interested in the quantity

\begin{equation}
\label{quarantasette}
W(\gamma)={1\over N} {\bf \Big< 0}|Tr{\cal T}{\cal P}\Big( exp\Big[ig
\oint_{\gamma}
A dx^+\Big]\Big)|{\bf 0 \Big>},
\end{equation}

where ${\cal T}$ means time-ordering and
${\cal P}$ color path-ordering along $\gamma$.

The vacuum state belongs to the physical Hilbert space as far as
the non vanishing frequency parts are concerned; it is indeed the
Fock vacuum $|{\bf \Omega \Big>}$. Then we
consider its direct product with the lowest eigenstate of the
Hamiltonian concerning zero modes (see \cite{BGN96}).
Due to the occurrence of zero modes, we cannot define a ``bona fide"
complete propagator for our theory:
on the other hand a propagator is not required in
eq.(\ref{quarantasette}).

We shall first discuss
the simpler case of QED, where no color ordering is involved.
Eq.(\ref{quarantasette}) then becomes

\begin{equation}
\label{quarantotto}
W(\gamma)= {\bf \Big< 0}|{\cal T}\Big( exp\Big[ig\oint_{\gamma}
A dx^+\Big]\Big)|{\bf 0 \Big>},
\end{equation}

and a little thought is enough to realize the factorization property
\begin{eqnarray}
\label{quarantanove}
W(\gamma)&=& {\bf \Big< 0}|{\cal T}\Big( exp\Big[{ig \over {\sqrt
{2L}}}\oint_{\gamma}
(b_0 + a_0t) dx^+\Big]\Big)|{\bf 0 \Big>}{\bf \Big< 0}|{\cal T}
\Big( exp\Big[ig \oint_{\gamma}
\hat{A}(t,x) dx^+\Big]\Big)|{\bf 0 \Big>}\nonumber\\
&=& W_0\cdot\hat{W},
\end{eqnarray}
according to the splitting of the potential in zero mode and frequency parts.
In turn the Wilson loop $\hat{W}$ can also be expressed as a Feynman
integral starting
from the QED lagrangian , without the zero mode

\begin{equation}
\label{cinquanta}
\hat{W}(\gamma)={\cal N}^{-1} \Big( exp\Big[g\oint_{\gamma}{\partial
\over {\partial J}}dx^+\Big]\Big)
\Big[\int {\cal D} \hat{A}\, {\cal D}\lambda\, exp\,\,i\Big(\int d^2x({\cal L}+
J\hat{A})\Big)\Big]_{\Big| J=0},
\end{equation}

${\cal N}$ being a suitable normalization factor.

Standard functional integration gives

\begin{eqnarray}
\label{cinquantuno}
\hat{W}(\gamma)&=&{\cal N}^{-1}\Big(exp\Big[g\oint_{\gamma}{\partial
\over {\partial J}}dx^+\Big]\Big) \nonumber\\
&& exp\Big[{i\over 2}\,\int\!\!\!\int d^2\xi d^2\eta
J(\xi) G_{c}(\xi-\eta)
J(\eta)\Big]_{\Big| J=0}.
\end{eqnarray}

and we are led to the expression

\begin{eqnarray}
\label{cinquantasette}
\hat{W}(\gamma)&=&exp\Big[ i\,g^2 \oint_{\gamma}dx^+\,\oint_{\gamma}
dy^+\,{G}_c(x^+-y^+,x^-\,-y^-)\Big]\nonumber\\
&=&exp\Big[ i\,g^2 \oint_{\gamma}dx^+\,\oint_{\gamma}
dy^+\,{G}(x^+-y^+,x^-\,-y^-)\Big]\nonumber\\
&=& exp\Big[-i\,{g^2\,{\cal A} \over 2}\Big]\,
exp\Big[ -i\,g^2 \oint_{\gamma}dx^+\,\oint_{\gamma}
dy^+\,{|x^+ + x^- -y^+ - y^-|\over {4L \sqrt 2}}\Big]
\end{eqnarray}

the absorptive part of the ``potential" averaging to zero in the abelian
case. Therefore the abelian Wilson-loop calculation is unable to
discriminate between the two different Green functions ${G}_c$ and
${G}$. The quantity ${\cal A}=\lambda \tau$ is the area of the loop.
The same result can also be obtained by operatorial
techniques, using Wick's theorem and the canonical algebra.

We are thereby left with the problem of computing $W_0$. In \cite{BGN96}
we have shown that the zero mode contribution
{\it exactly cancels} the last exponential in eq.(\ref{cinquantasette}),
leaving the pure loop area result, only as a consequence of the
canonical algebra, and even in the presence of
a topological degree of freedom. The result  coincides
with the one we would have obtained introducing in eq.
(\ref{cinquantasette}) the complete Green's function, $i.e.$ with
the zero mode included. The same area result is obtained also in the non
abelian case if we use the 't Hooft form for the propagator
${1\over 2}|t|\delta_{p}(x+t)$, in spite of
the fact that this form has not a sound canonical basis and that
factorization in (\ref{quarantanove}) is no longer
justified in the non abelian case. As a matter of fact a little thought
is enough to realize that only planar diagrams survive thanks to the
`contact' nature of the potential, leading to the expression

\begin{equation}
\label{eresia3}
W(\gamma)= exp\Big[-i\,{g^2\,C_F\,\lambda\,\tau \over 2}\Big].
\end{equation}

The area (${\cal A}=\lambda\,\tau$) law behaviour of
the Wilson loop we have found in this case together with the occurrence
of a simple exponentiation in terms
of the Casimir of the fundamental representation, is a quite
peculiar result, insensitive to the decompactification limit
$L\to \infty$. It is rooted in the particularly simple expression
for the ``potential" we have used that coincides with the one
often considered in analogous Euclidean calculations \cite{BR80}.

However canonical quantization suggests that we should rather
use the propagator $G_{c}(t,x)$.
A full resummation of perturbative exchanges is
no longer viable in this case,
owing to the presence of non vanishing cross diagrams, in which
topological excitations mix non trivially with the frequency parts.
Already at ${\cal O}(g^4)$, a tedious but straightforward calculation
of the sum of all the ``cross" diagrams  leads to the result

\begin{equation}
\label{follia1}
W_{cr}= -({g^2 \over {4\pi}})^2\, 4\,C_F\, (C_F - {{C_A}\over 2})
({\cal A})^2 \int_{0}^{1}
d\,\xi \int_{0}^{1} d\,\eta\,log{|sin\rho(\xi-\eta)|\over
{|sin\rho\xi|}}log{|sin\rho(\xi-\eta)|\over
{|sin\rho\eta|}},
\end{equation}

where $\rho= {\pi\lambda\over{\sqrt 2 L}}$.
One immediately recognizes the appearance of the
quadratic Casimir of the adjoint
representation ($C_A$); moreover,  a
dimensionless parameter $\rho$, which measures the ratio of the side
length $\lambda$ to the interval length $L$, explicitly occurs.
In the decompactification limit $\rho\to 0$, the expression of the
cross graph
given in ref. \cite{BDG94}

\begin{equation}
\label{follia2}
W_{cr}=-({g^2 \over {4\pi}})^2\,2 C_F\,(C_F -C_A/2)
 ({\cal A})^2 {\pi^2\over 3}
\end{equation}

is smoothly recovered. Of course the finite self-energy contribution
found in ref. \cite{BDG94} in the dimensionally regularized theory
cannot appear in a strictly 1+1 dimensional treatment.
It is perhaps not surprising that in the
limit $L \to \infty$
the perturbative result for $W_{cr}$ in the continuum is correctly
reproduced, in spite
of the presence of topological excitations.
Still the difference between the result obtained with the `contact'
potential and with the causal one is even more striking: in both cases
at large N only planar diagrams survive but they nevertheless give
rise to different expressions for the same Wilson loop.

It seems that $planarity$ is not enough to single out an unambiguous
result.

\section{The 't Hooft bound state equation}
\noindent
In 1974 G. 't Hooft \cite{TH74} proposed a very interesting model to describe
the mesons, starting from a SU(N) Yang-Mills theory in 1+1 dimensions
in the large N limit.

Quite remarkably in this model quarks look confined, while a discrete
set of quark-antiquark bound states emerges, with squared masses lying
on rising Regge trajectories.

The model is solvable thanks to the ``instantaneous'' character of
the potential acting between quark and antiquark.

Three years later such an approach was criticized by T.T. Wu \cite{WU77},
who replaced the instantaneous 't Hooft's potential by an expression
with milder analytical properties, allowing for a Wick's rotation
without extra terms.

Unfortunately this modified formulation led to a quite involved bound
state equation, which may not be solved. An attempt to treat it
numerically in the zero bare mass case for quarks \cite{BS78} led only to
partial answers in the form of a completely different physical
scenario. In particular no rising Regge trajectories were found.

After those pioneering investigations, many interesting papers
followed 't Hooft's approach, pointing out further remarkable
properties of his theory and blooming into the recent achievements
of two dimensional QCD , whereas Wu's approach sank into oblivion.

Still, equal time canonical quantization of Yang-Mills theories
in light-cone gauge \cite{BDLS85} leads precisely in 1+1 dimensions
to the Wu's expression for the
vector exchange between quarks \cite{BDG94}, which is nothing but the 1+1
dimensional version of the Mandelstam-Leibbrandt (ML)
propagator. We have already stressed that this option is mandatory
in order to achieve gauge
invariance and renormalization in 1+(D-1) dimensions.

We follow here definitions and notations of
refs.\cite{TH74} and \cite{WU77} the reader is
invited to consult.

The 't Hooft potential exhibits an infrared singularity
which, in the original formulation, was handled by introducing
an infrared cutoff; a quite remarkable feature of this theory
is that bound state wave functions and related eigenvalues
turn out to be cutoff independent. As a matter of fact in
ref. \cite{CA76}, it has been pointed out that the singularity
at $k_{-}=0$ can also be regularized by a Cauchy principal
value ($CPV$) prescription without finding differences in gauge
invariant quantities. Then, the difference between the two
potentials is represented by the following distribution

\begin{equation}
\label{unoa}
\Delta (k)\equiv {{1}\over {(k_{-}-i\epsilon
sign (k_{+}))^2}} - P\Big({{1}\over {k_{-}^2}}\Big)= - i \pi
sign (k_{+}) \delta^{\prime}(k_{-}).
\end{equation}

In ref.\cite{BG96}, which we closely follow in the sequel,
this quantity has been treated as an insertion
in the Wu's
integral
equations for the quark propagator and for the bound state wave
function, starting from 't Hooft's solutions. $Exactly$
the same planar diagrams of
refs.\cite{TH74} and \cite{WU77}, which are the relevant ones
in the large $N$ limit, are summed.

The Wu's integral equation for the quark self-energy in the Minkowski
momentum space is

\begin{eqnarray}
\label{unob}
\Sigma(p;\eta)&=& i {{g^2}\over {\pi^2}} {{\partial}\over {\partial p_{-}}}
\int dk_{+}dk_{-} \Big[P\Big({{1}\over {k_{-}-p_{-}}}\Big)+
i \eta \pi sign (k_{+}-p_{+}) \delta (k_{-}-p_{-})\Big]\nonumber\\
&\cdot&{{k_{-}}\over {k^2+m^2-k_{-}\Sigma (k;\eta)-i\epsilon}},
\end{eqnarray}

where $g^2=g_0^2 \,N$ and $\eta$ is a real parameter which is used
as a counter of insertions and eventually should be set equal to 1.

Its exact solution with appropriate boundary conditions reads

\begin{eqnarray}
\label{uno}
\Sigma(p;\eta)&=& {{1}\over {2p_{-}}}\Big(\Big[p^2+m^2+(1-\eta){{g^2}\over
{\pi}}\Big]-\Big[p^2+m^2-(1-\eta){{g^2}\over
{\pi}}\Big]\nonumber\\
&\cdot&\sqrt {1- {{4\eta g^2 p^2}\over {\pi(p^2+m^2-(1-\eta){{g^2}\over
{\pi}}}-i\epsilon)^2}}\,\,\Big).
\end{eqnarray}

One can immediately realize that 't Hooft's and Wu's solutions
are recovered for $\eta =0$ and $\eta =1$ respectively.

The dressed quark propagator turns out to be

\begin{equation}
\label{due}
S(p;\eta) = - {{i p_{-}}\over {m^2+2 p_{+}p_{-}- p_{-}\Sigma(p;\eta)}}.
\end{equation}

Wu's bound state equation in
Minkowski space, using light-cone coordinates, is

\begin{eqnarray}
\label{tre}
\psi(p,r)&=& {{-ig^2}\over {\pi ^2}} S(p;\eta) S(p-r;\eta)
\int dk_{+}dk_{-} \Big[P\Big({{1}\over
{(k_{-}-p_{-})^2}}\Big)-\nonumber\\
&-&
i \eta \pi sign (k_{+}-p_{+}) \delta^{\prime} (k_{-}-p_{-})\Big]
\psi(k,r).
\end{eqnarray}

We are here considering for simplicity the equal mass case and $\eta$
should be set equal to 1.

Let us denote by $\phi_{k}(x),\,\, 0\le \! x= {{p_{-}}\over {r_{-}}}\le
\!1,\,\, r_{-}>0$,
the 't Hooft's eigenfunction corresponding
to the eigenvalue $\alpha_{k}$ for the quantity ${{-2 r_{+}r_{-}}\over
{M^2}}$, where $M^2= m^2 - {{g^2}\over {\pi}}$.
Those eigenfunctions are
real, of definite parity under the exchange $x \to 1-x$ and vanishing
outside the interval $0<x<1$:

\begin{eqnarray}
\label{quattro}
\phi_{k}(x)&=& \int dp_{+} {{r_{-}}\over {M^2}}
\psi_{k}(p_{+},p_{-},r),\nonumber\\
i \pi\,\psi_{k}&=&\phi_{k}(x) {{M^4}\over {M^2+2r_{-}p_{+}x
-i\epsilon}}\cdot\nonumber\\
&\cdot&{{1-\alpha_{k}x(1-x)}\over {M^2-\alpha_{k}M^2(1-x)-
2r_{-}p_{+}(1-x)- i \epsilon}}.
\end{eqnarray}

They  are solutions of eq.(\ref{tre}) for $\eta=0$ and
form a complete set.

We are interested in a first order calculation in $\eta$.
This procedure
is likely to be sensible only in the weak coupling region
${{g_0^2}\over {\pi}}< m^2$.
The integral equation (\ref{tre}), after first order
expansion in $\eta$ of its kernel, becomes

\begin{eqnarray}
\label{cinque}
\psi(p_{+},p_{-},r)&=& {{ig^2}\over {\pi^2}}{{p_{-}}\over
{M^2+2p_{+}p_{-}-i\epsilon}}{{p_{-}-r_{-}}\over
{M^2+2(p_{+}-r_{+})(p_{-}-r_{-})-i\epsilon}}\nonumber\\
\cdot\Big[\Big(1-{{\eta g^2 M^2}\over
{\pi}}&[&(M^2+2p_{+}p_{-}-i\epsilon)^{-2}+(M^2+2(p_{+}-r_{+})(p_{-}-r_{-})
-i\epsilon)^{-2}]\Big)\nonumber\\
&\cdot&\int dk_{+}dk_{-}
P{{1}\over{(k_{-}-p_{-})^2}}\psi(k_{+},k_{-},r)-\nonumber\\
&-&i \pi \eta \int dk_{+}dk_{-} sign(k_{+}-p_{+})
\delta^{\prime}(k_{-}-p_{-})\psi(k_{+},k_{-},r)\Big].
\end{eqnarray}

We integrate this equation over $p_{+}$ with $r_{-}>0$ and look
for solutions with the same support properties of 't Hooft's ones.
We get

\begin{eqnarray}
\label{cinquea}
&&\phi(x,r)= {{g^2}\over{\pi M^2}}{{x(1-x)}\over{1-\alpha
x(1-x)-i\epsilon}}\Big[\Big(1-\eta{{g^2}\over{\pi M^2}}
{{x^2+(1-x)^2}\over{(1-\alpha
x(1-x)-i\epsilon)^2}}\Big)\nonumber\\
\cdot &P&\int_0^1 {{dy}\over {(y-x)^2}}\phi(y,r)
-{{\alpha \eta}\over{2}}\int d\xi \log {{{{1}\over{1-x}}-\alpha (1-\xi)-i
\epsilon}\over {{{1}\over{x}}-\alpha
\xi-i\epsilon}}\psi^{\prime}(\xi,x,r)\Big],
\end{eqnarray}

where $\prime$ means derivative with respect to $x$.

It is now straightforward to check that 't Hooft's solution
$\psi_{k}(p_{+},p_{-},r)$ is indeed a solution also of this
equation when $\alpha$
is set equal to $\alpha_{k}$, for any value of $\eta$, in particular
for $\eta=1$, thanks to a precise cancellation of the contributions
coming from the propagators (``virtual'' insertions) against the
extra term due to the modified form of the ``potential'' (``real''
insertion). In other words the extra piece of the kernel at
$\alpha=\alpha_{k}$ vanishes when acting on $\psi_{k}$ as a perturbation. This
phenomenon is analogous to the one occurring, with respect to the same extra
term, in one loop perturbative four-dimensional  calculations concerning
Altarelli-Parisi  \cite{BA93} and Balitsky-Fadin-Kuraev-Lipatov
\cite{BR93} kernels. This analogy may have far-reaching consequences.

As a matter of fact, taking 't Hooft's equation
into account, we get

\begin{eqnarray}
\label{sei}
&&\Big[1-{{\eta g^2}\over{\pi M^2
[1-\alpha x(1-x)-i\epsilon]^2}}\Big((1-x)^2+
+[x^2[1+{{1-\alpha x(1-x)}\over{1-\alpha_{k}
x(1-x)-i\epsilon}}]\Big)\Big]\cdot\nonumber\\
&\cdot&(\alpha_{k}-\alpha)\phi_{k}(x)=
{{\eta g^2}\over {\pi
M^2}}\,\,\phi_{k}^{\prime}(x)\,\, log {{1-\alpha_{k}x(1-x)-i\epsilon}
\over{1-\alpha x(1-x)-i\epsilon}}.
\end{eqnarray}

There are no corrections from a single insertion in the kernel to
't Hooft eigenvalues and eigenfunctions.
We stress that this result does
not depend on their detailed form,
but only on their general properties.
The ghosts which are responsible of the causal
behaviour of the ML propagator do not modify
the bound state spectrum, as their ``real'' contribution
cancels against the ``virtual'' one in propagators.
Wu's equation for colorless bound states,
although much more involved than the
corresponding 't Hooft's one, might still apply.
This is
the heuristic lesson one learns from a single insertion in the kernel
and is in agreement with the mentioned similar mechanism occurring
in four-dimensional perturbative QCD.

Unfortunately this conclusion holds only at the level of a single
insertion and may be a consequence of one loop unitarity which
tightly relates `real' to `virtual' exchanges. Already when
two insertions are taken into account, deviations are seen from
't Hooft spectrum \cite{BNS96}. This is not a surprise as
Wu's equation is deeply
different from 't Hooft's one and might describe
the theory in a different phase (see for instance
\cite{ZY95}).

Planarity plays a crucial role in both formulations; indeed the two equations
sum exactly the same set of diagrams (the planar ones), which are thought to
be the most important ones in the large N limit. The first lesson one
learns is that planarity by itself is not sufficient to set up unambiguously
a physical scenario.

Now there are good arguments \cite{ZY95} explaining why
planarity should break down in the limit $m\to 0$. The same situation
should occur when $m^2<{{g_0^2}\over \pi}$ which correspond to a `strong'
coupling situation, where we know that 't Hooft solution can no longer
be trusted.

What about the `weak' coupling regime? Should we believe that
't Hooft's picture describes correctly the physics in two
dimensions, which in turn should be represented by planar diagrams,
we would conclude that in 1+1 dimensions planarity is not a good approximation
in the causal formulation of the theory. Indeed the results we obtain
in the latter case are definitely different from 't Hooft ones.
This is a very basic issue
in our opinion, which definitely deserves further investigation.
This is even more compelling should this situation persist in
higher dimensions where causality is mandatory in order to
obtain an acceptable formulation of the theory.
\vskip .5truecm

\nonumsection{Acknowledgements}
\noindent
We thank L. Griguolo for many useful discussions and friendly
collaboration.

\nonumsection{References}

\end{document}

\bye